\documentstyle[aps,version2]{revtex}

\begin{document}
\pagestyle{empty}

\begin{title}
Spin-dependent structure functions $\hat g_1$ and $\hat g_2$ for
inclusive spin-half baryon production in electron-positron annihilation
\end{title}

\author{Wei Lu$^{1,2}$, Xueqian Li$^{1,3}$ and Haiming Hu$^4$}

\begin{instit}
$1$. CCAST (World Laboratory), P.O. Box 8730, Beijing 100080, China

$2$. Institute of High Energy Physics,
 P.O. Box 918(4), Beijing 100039, China\footnote{Mailing address}

$3$. Department of Physics, Nankai University,
Tianjin 300071, China

$4$. Department of Physics, Peking University,
Beijing 100871, 300071, China
 \end{instit}

\begin{abstract}
 Two spin-dependent structure functions
$\hat g_1$ and $\hat g_2$ for the inclusive
spin-half baryon production in electron-positron annihilation
are studied in the context of QCD factorization as well as
in the naive quark parton model.  As a result,  it is found
that  the sum of $\hat g_1$ and $\hat g_2$ is related to
$\hat h_1$ and $\hat g_T$, two quark fragmentation functions
defined by Jaffe and Ji.  In connection with
the  measurement of quark fragmentation functions,
the possible phenomenological consequences are discussed.
\end {abstract}


\newpage
\pagestyle{plain}
\small

\section{Introduction}

In the development of our understanding of the hadron
production mechanism, the inclusive one-particle production
 by electron-positron annihilation plays the same role
as the deeply inelastic scattering does in our learning
of the nucleon structures.  Experimentally, a number of
electron-positron colliders have been operated and
will be  constructed at various center of mass energies.
On the theoretical side,
a series of parton fragmentation functions
have  been identified by Jaffe and Ji \cite{JJ,Ji}, which
characterize the information about the inclusive  hadron production.
Therefore, it is worthwhile to explore the possibility  to
measure them, expecially those spin-dependent ones,  by
the inclusive one-particle production in electron-positron
annihilation.
To study spin physics  related to the inclusive spin-half
baryon production,  it is usually needed to monitor the
polarization of the produced particles. If  we are restricted with
the inclusive hyperon production,  it is  quite feasible
to measure the polarization of the produced particles.
In fact, a series of experiments
have already  been done
at Fermilab measuring the polarization of the inclusive hyperons
in fixed-target experiments \cite{hyperon}.  We also note that some preliminary
results \cite{pre} on the measurement of the $\Lambda$ polarization
at CERN  LEP  have been existent as well.   Because what
can be accessed directly  by experiments  are structure
functions,  we  explore in this paper
the relation between $\hat g_1$ and $\hat g_2$,
two spin-dependent structure functions
for the inclusive spin-half baryon
production by electron-positron annihilation, and
the quark fragmentation functions defined by Jaffe and Ji.

 Without loss of any  generalities,  we consider  the
inclusive $\Lambda$ hyperon  production
for among the hyperons the $\Lambda$ particle
is of the largest production cross section and
its polarization is the easiest  to measure. The advantages
of measuring  the polarization of the inclusive $\Lambda$
hyperon  are also realized by several groups  of authors
in different cases. Among others, Burkardt and Jaffe
\cite{cite1} investigated the possibility to measure
the $s\to \Lambda$ fragmentation functions at LEP,
and Chen {\it et. al.}  \cite{cite2} discussed the  more complicated
semi-inclusive process $e^+ e^- \to \Lambda X\bar \Lambda \bar X$.

For simplicity,
we work in the energy region not too high so that
only the photon channel needs to be considered.
Then, all the information about the hyperon production
is entailed by the photon fragmentation tensor, which is
defined as
\begin{equation}
\hat W_{\mu\nu}(q,p,s)=\frac{1}{4 \pi}
\sum \limits_X \int d^4 \xi \exp ( iq \cdot \xi)
\langle0| j_\mu (0)|\Lambda (p,s),X\rangle
\langle\Lambda (p,s),X|j_\nu (\xi)|0\rangle,
\end{equation}
where $\sum \limits_X$ represents the summation over all the possible
final states that contain the inclusive $\Lambda$  hyperon.
Throughout the  work, we  normalize the spin vector
in such a way that $s\cdot s =-1$ for a pure state.  The
electromagnetic  current is defined as $j^\mu=\sum\limits_f
e_f \bar \psi_f \gamma^\mu \psi_f$, with $f$ being the
quark flavor index and $e_f$ being the electric charge of the
quark in unit of the electron charge.  In our presentation,
we will suppress the flavor index whenever possible.
Subjected to the gauge invariance,  hermiticity
and parity conservation, $\hat W(q,p,s)$  assumes the following
general Lorentz decomposition \cite{lw}:
\begin{eqnarray}
\label{dec}
\hat W_{\mu\nu}(q,p,s)&
=&\displaystyle\frac{1}{2}
\left[ (-g_{\mu\nu}+\frac{q_\mu q_\nu}{q^2})\hat F_1(z_B,Q^2)
+(p_\mu -\frac{p\cdot q}{q^2}q_\mu)
(p_\nu -
\frac{p\cdot q}{q^2}q_\nu)
\frac{\hat F_2(z_B,Q^2)}{p\cdot q} \right]
\nonumber  \\
& &
+iM \varepsilon_{\mu\nu\lambda\sigma}q^\lambda s^\sigma
\frac{\hat g_1(z_B,Q^2)}{p\cdot q}
+iM \varepsilon_{\mu\nu\lambda\sigma}q^\lambda
(s^\sigma -\frac{s\cdot q}{p\cdot q} p^\sigma)
\frac{\hat g_2(z_B,Q^2)}
{p\cdot q} \nonumber \\
& &
+M \left[(p_\mu-\displaystyle\frac{p\cdot q}{q^2}
q_\mu)\varepsilon_{\nu \rho\tau\eta}p^\rho q^\tau s^\eta
+(p_\nu-\displaystyle\frac{p\cdot q}{q^2}q_\nu)
\varepsilon_{\mu \rho\tau\eta}p^\rho q^\tau s^\eta
\right ]\frac{\tilde F
(z_B,Q^2)}{(p\cdot q)^2},
\end{eqnarray}
where  $z_B\equiv 2 (p\cdot q)/q^2$, $Q\equiv \sqrt{q^2}$,
$M$ is the mass of the  $\Lambda$ hyperon,
and $\hat F_1$, $\hat F_2$, $\hat g_1$,
$\hat g_2$, and $\tilde F$ are the
scaling structure functions. It should be noted
that due to the final-state interactions, the time
reversal invariance  cannot exert any constraints on the
Lorentz decomposition of $\hat W^{\mu\nu}(q,p,s)$.

Obviously, there exists one-to-one correspondence
between $\hat F_1$, $\hat F_2$, $\hat g_1$,
and $\hat g_2$, on the one hand, and four structure
functions $F_1$, $F_2$, $g_1$ and $g_2$ in the deeply inelastic
scattering, on the other hand.
$\hat F_1$ and $\hat F_2$ have been  vastly
discussed in the literature \cite{webber}, so we will simply ignore them
in our discussion.  As for the structure function  $\tilde F$, which
 arises from the  final-state interaction,  we
will not address it, either, to avoid dispersing the reader's sight.
Our attention  will be focused on
$\hat g_1$ and $\hat g_2$  so that our results
can be compared with their counterparts in
the deeply inelastic scattering case \cite{dis}.

We establish our
coordinate system by letting the $\hat z$ axis
be along the outgoing direction of the inclusive
hyperon and the $\hat x$--$\hat z$ plane in the
production plane. We adopt the light-cone coordinates
and parameterize the $\Lambda$ momentum as
\begin{equation}
p^\mu=P^\mu +\frac{1}{{2}} M^2 n^\mu,
\end{equation}
where
\begin{equation}
P^\mu =\frac{1}{\sqrt {2}} ( \sqrt{M^2 +|{\bf p}|^2} +|{\bf p}|)
 (1^+, 0^-, 0_\perp),
\end{equation}
\begin{equation}
 n^\mu =\frac{\sqrt{2}}{ M^2}( \sqrt{M^2 +|{\bf p}|^2} -|{\bf p}|)
(0^+, 1^-, 0_\perp).
\label{n}
\end{equation}
with $|{\bf p}|$ being the  magnitude of the $\Lambda$ momentum.
We  will work in the frame in which  $|{\bf p}|$  has a  large value.
Obviously,  $P$ and $n$ are light-like and  they satisfy $P\cdot  n =1$.

\section{Naive Parton Model Results}

 We feel it desirable to
present a naive parton model approach to $\hat g_1$ and $\hat g_2$,
since  the  corresponding results  can supply us with  a
benchmark  for comparison.  In the naive parton model,
the fragmented hadrons  will be completely collinear to their
parent quark parton, with no transverse momentum effects.
Correspondingly,  we need to assign a mass $m_f=M/z$ to
the quark of flavor $f$, where $z$ is the momentum
fraction of the quark carried by the inclusive $\Lambda$ particle.
$\hat W^{\mu\nu} (q,p,s)$  can be obtained from
the convolution of the underlying
tensor $\hat w^{\mu\nu}_f(q,z,s_f)$ for photon fragmentation
into a quark-antiquark pair with the $\Lambda$ multiplicity
distribution  $N_{f\to \Lambda}(z, s_f, s)$ in the jet initiated
by the quark of flavor $f$ and spin four-vector $s_f$:
\begin{equation}
\hat W^{\mu\nu} (q,p,s)=\sum\limits_f \int  \frac{dz}{z}
\delta (z-z_B) \hat w^{\mu\nu}_f(q,z,s_f) N_{f\to \Lambda}(z, s_f, s).
\end{equation}
By analogy with the well-known  leptonic tensor, we know
\begin{equation}
\hat w^{\mu\nu}_f(q,z,s_f) =e^2_f \left[
(-g^{\mu\nu}+ \frac{q^\mu q^\nu}{q^2})
q^2
 -\frac{4}{z^2}
(p^\mu - \frac{p\cdot q}{q^2}q^\mu)
(p^\nu - \frac{p\cdot q}{q^2}q^\nu)
+2im_f \epsilon^{\mu\nu\tau\rho}q_\tau s_{f\rho}\right].
\end{equation}
Therefore, we  have the following results for $\hat g_1$ and $\hat g_2$:
\begin{equation}
\hat g_1(z_B)= \frac{1}{z_B}\sum\limits_f e^2_f \Delta  D_f (z_B, s),
\end{equation}
\begin{equation}
\hat g_2(z_B)=0,
\end{equation}
where
\begin{equation}
\Delta  D_f (z_B, s)= N_{f\to \Lambda}(z, s, s)
-N_{f\to \Lambda}(z, -s, s)
\end{equation}
is the $\Lambda$ number density difference with
spin parallel and antiparallel to the quark spin  in the $f$-flavored
quark jet. In the naive parton model, if one boosts into the
quark rest frame, all its fragmenting products will be at rest, too.
Hence, $\Delta  D_f (z_B, s)$  does not depend on the spin
quantization direction \cite{parton}.

\section{QCD Factorization Approach}

In the QCD field theory approach,
the photon  fragmentation tensor is composed of the contributions
of an infinite series of the  so-called cut Feynman diagrams.
The cut diagram
is formed by piecing together the Feynman diagram  of one amplitude
with  the complex conjugate  of another one  with  the same
initial and final states.  Owing to lack of the
methods to treat nonperturbative interactions, one
usually adopt the strategy to factorize \cite{fact}
 the hadron-involved process into short- and long-distance parts.
The former can be calculated perturbatively whereas
the latter can be measured experimentally.
In  such a factorization approach,
QCD long-distance interactions are
conventionally represented by blobs in the cut diagram.
In  general, these cut diagrams
begin to contribute at the different powers of the hard
scattering scale.  If a  physical
quantity goes like $Q^{\tau-2}$ or begins to
make contributions at $O(Q^{\tau -2})$,  it is referred to as
being  twist-$\tau$.  According to Qiu and Sterman \cite{QS-fac2},
the twist at which a cut diagram begins to contribute is
controlled by the number of $physical$ parton lines that
attach to the long-distance interaction blobs.
Now it is well known that  some potentially power-suppressed
twist-three spin asymmetries  are relatively large,  so in this paper
we will work up to the twist-three level, i.e., to the
first nonleading  power. In addition, we work  in the light-cone
gauge specified by  $n\cdot A=0$.
Correspondingly,   the  cut diagrams shown  in figs. 1, and 2
consist of  our subjects.

\subsection{Longitudinally-polarized $\Lambda$ hyperon}

   We begin  with the case  in which  the
inclusive $\Lambda$  is  of plus helicity.
In this case, the leading  spin dependence
of  $\hat W^{\mu\nu} (q,p,s)$ comes about at
twist two, while  its  next-to-leading spin
dependence  at twist four.  To protrude the
main physics, we work at leading twist and
accordingly only the lowest-order diagram
shown in fig. 1 gets into work. As a result,
\begin{equation}
\hat W^{\mu\nu}(q,p,s_{||})= \frac{1}{4\pi N}
\int \frac{d^4 k}{(2\pi)^4}
{\rm Tr_D} {\rm Tr_C} \left[
H^{\mu\nu}_{(1)}(q,k)T(k,p,s_{||})\right],  \label{t2}
\end{equation}
where ${\rm Tr_D}$ and ${\rm Tr_C}$ stands for making traces in
the  Dirac and color spaces, and
\begin{equation}
T_{\alpha\beta}(k,p,s_{||})=
\sum\limits_X\int d^4 \xi \exp(-ik\cdot \xi)
\langle0|\psi_{\alpha}(0)|\Lambda(p,s_{||}),X\rangle\langle\Lambda(p,s_{||}),X)
|\bar \psi_{\beta}(\xi)|0\rangle.
\end{equation}
Throughout all, we  reserve the subscripts $_\alpha$ and $_\beta$
for the Dirac indices.  The color summation is
implicit in our parton fragmentation matrices,  so
there is a  color factor $1/N$ ($N=3$) on the right-hand side
of Eq. (\ref{t2}).
Since we work in the physical gauge,  the
gauge linkages of the form
$ {\rm P} \exp [-i g \int^\infty_y dy^{\prime\mu}
A^\mu (y^\prime)]$ are identical to unity
and hence suppressed in our parton fragmentation matrices.

 Since high-energy  reactions are light-cone dominant,
the  most efficient way to extract the leading contributions
from each diagram is to  make an expansion about the
components of  the parton momenta  that are collinear
to the corresponding hadron momentum.  After that,
one must further decouple the  discrete indices
between the long-distance matrices and the  associated
short-distance parts.  For completing these tasks,
one can employ the collinear
technique developed by Ellis, Furmanski and Petonzio \cite{EFP}.

 Let us parameterize the quark momentum as
\begin{equation}
k^\mu=\frac{1}{z} P^\mu +k^\mu_T +\frac{k^2-k^2_T}{2 k\cdot n} n^\mu.
\end{equation}
Then, the collinear expansion for the lowest-order diagram reads
\begin{equation}
H_{\mu\nu} (q,k)= H_{\mu\nu} (q,P/z)
+ \frac{\partial H_{\mu\nu} (q,k) }{\partial  k^\sigma}
|_{k=P/z} (k-P/z)^\sigma
+\cdots~ .
\label{B-ex}
\end{equation}
Since working  at twist two, here we  need only to consider the
the contributions associated with the leading term.
As  a result, we obtain
\begin{equation}
\hat W^{\mu\nu}(q,p,s_{||})= \frac{1}{4\pi N}
\int \frac{dz}{z}
{\rm Tr_D} {\rm Tr_C} \left[
H^{\mu\nu}_{(1)}(q,P/z)T(z,p,s_{||})\right],
\end{equation}
with
\begin{equation}
T_{\alpha\beta} (z,p,s_{||})=
z\sum\limits_X\int  \frac{d\lambda }{2 \pi}
\exp ( -i\lambda/z)
\langle0|\psi_{\alpha} (0)|\Lambda (p,s_{||}),X\rangle
\langle\Lambda (p,s_{||}),X|\bar \psi_{\beta} (\lambda n )|0\rangle.
\label{xm1}
\end{equation}
Since the color index has  been decoupled in Eq. (\ref{t2}),
now the only task for us is to decouple the Dirac indices between the
expanded hard-interaction part $H^{\mu\nu}_{(1)}(q,P/z)$ and the simplified
quark fragmentation matrix $T(z,p,s_{||})$.  Expanding the latter
in the Dirac space, we have
\begin{equation}
T(z,p,s_{||})= M(s_{||}\cdot n)\gamma_5 \rlap/P \hat {\sl g}_1 (z)
+M (s_{||}\cdot n)(\rlap/P \rlap/n-
\rlap/n \rlap/P) \gamma_5\hat  h_L (z)
+M^2 (s_{||}\cdot n)\rlap/n \gamma_5 \hat {\sl g}_3 (z)
+\cdots,  \label{dc}
\end{equation}
where we have suppressed the spin-independent terms as well as
those arising from the hadronic final-state interaction
because we do not address $F_1$ and $F_2$ as well as $\tilde F$.
$\hat {\sl g}_1 (z)$, $\hat h_L (z)$ and $\hat {\sl g}_3 (z)$
are quark fragmentation functions (matrix elements)
at twist two, three, and four respectively, and their definitions
can be found in Ref. \cite{Ji}.  From the chiral-odd
structure of $\hat h_L (z)$, one can immediately see that
only by a quark mass insertion in the hard part can it make
a  nonzero contribution, so its contribution is actually at twist four.

 Keeping the first term in Eq. (\ref{dc}), we will have
\begin{equation}
\hat W^{\mu\nu}(q,p,s_{||})= \frac{i}{z_0 P\cdot q}
 \epsilon^{\mu\nu\tau\rho}
q_\tau s_{|| \rho} \sum\limits_f e^2_f \hat {\sl g}_1 (z_0),
\end{equation}
where $z_0\equiv 2 (P\cdot q)/q^2$.  Notice that we
have recovered the quark flavor index here. To confront with
the general decomposition of  $W^{\mu\nu}(q,p,s_{||})$,
we can simply substitute $p$ for $P$ and correspondingly
$z_B$ for $z_0$. Such replacements amount to including
some effects at  two higher twist so they are allowed.
As a consequence, we  have the following  twist-two formulas:
\begin{equation}
\hat g_1(x_B) =\frac{1}{z_B}\sum\limits_f e_f^2
\hat {\sl g}_1 (z_B),
\label{ss1}
\end{equation}

\begin{equation}
g_2(x_B)=0.
\end{equation}
In fact, our factorization at leading twist reproduced
the naive parton model results
about $\hat g_1$ and $\hat g_2$. The easiest way to recognize
this point is to  quantize the quark filed in the light-cone
quantization formalism \cite{Jxd} and carry out the integration
over the unobserved hadron system \cite{Mulders}.
 Physically,  if one  notices that
the quark transverse momentum  with respect to the $\Lambda$
momentum  has been integrated out,  it is not difficult to
understand  the equivalence of  this twist-two factorization
result to the naive parton model approach.

\subsection{Transversely-polarized $\Lambda$ hyperon}

Now we turn to the case in which  the spin of  the
inclusive  $\Lambda$ hyperon  is  aligned  along
the normal of the production plane,  namely, the
$\hat y$ axis.
In this case,  $s\cdot q$=0, so the
$\hat g_1$ and $\hat g_2$  terms in Eq. (\ref{dec})
 become degenerate.
Therefore, we hope to   express the sum of
$\hat g_1$ and $\hat g_2$   in terms of some
spin-dependent quark fragmentation functions.
Related to the transverse spin  of the inclusive $\Lambda$
particle, the twist-two quark fragmentation function
$\hat h_1$ is  chiral-odd.  As a result,
its contribution comes about via a quark mass insertion
in the hard part and correspondingly is at twist three.
Hence, we need to work at least at twist three
and the lowest-order diagram as well as those two in fig. 2
construct our subject.

To the order at which we work,
\begin{eqnarray}
\hat W^{\mu\nu}(q,p,s_\perp)&=& \frac{1}{4\pi N}
\int \frac{d^4 k}{(2\pi)^4}
{\rm Tr_D} {\rm Tr_C} \left[
H^{\mu\nu}_{(1)}(q,k)T(k,p,s_\perp)\right]
\nonumber \\
& &
+\frac{1}{4\pi N}\int \frac{d^4 k}{(2\pi)^4}
\frac{d^4 k_1}{(2\pi)^4}
{\rm Tr_D} {\rm Tr_C}\left[
H^{\mu\nu\sigma}_{(2a)}(q,k,k_1)
X^\prime _\sigma(k,k_1,p,s_\perp) \right]
\nonumber \\
& &
+\frac{1}{4\pi N}\int \frac{d^4 k}{(2\pi)^4}
\frac{d^4 k_1}{(2\pi)^4}
{\rm Tr_D} {\rm Tr_C}\left[
H^{\mu\nu\sigma}_{(2b)}(q,k,k_1)
Y^\prime_\sigma(k,k_1,p,s_\perp)\right],
\label{www}
\end{eqnarray}
where
\begin{equation}
T_{\alpha\beta}(k,p,s_\perp)=
\sum\limits_X\int d^4 \xi \exp(-ik\cdot \xi)
\langle0|\psi_{\alpha}(0)
|\Lambda(p,s_\perp),X\rangle\langle\Lambda(p,s_\perp),X)
|\bar \psi_{\beta}(\xi)|0\rangle,
\end{equation}
\begin{eqnarray}
X^{\prime \sigma}_{\alpha\beta} (k_1,k,p,s_\perp)&=&
\sum\limits_X\int d^4\xi d^4 \xi_1
\exp \left (-i(k-k_1)\cdot \xi_1 -i k\cdot \xi\right)
\nonumber  \\
&  & \times
\langle0|(-g)A^\sigma (\xi_1)\psi_\alpha(0)
|\Lambda (p,s_\perp),X\rangle\langle\Lambda (p,s_\perp),X|
\bar\psi_\beta(\xi n) |0\rangle,
\label{u1}
\end{eqnarray}
\begin{eqnarray}
Y^{\prime \sigma}_{\alpha\beta} (k_1,k,p,s_\perp)&=&
\sum\limits_X\int d^4\xi d^4 \xi_1
\exp \left(-i(k_1-k)\cdot \xi -i k_1\cdot \xi_1\right)
\nonumber  \\
&  &\times  \langle0|
\psi_\alpha(0)|\Lambda (p,s_\perp),X\rangle\langle\Lambda (p,s_\perp),X|
\bar\psi_\beta(\xi_1   n)(-g)A^\sigma (\xi) |0\rangle.
\label{v1}\end{eqnarray}
Here we note that in writing down Eq. (\ref{www}),   we have
absorbed the color matrix along with a minus strong coupling $-gT^a_{ij}$
into $X^{\prime\sigma}_{\alpha\beta} (k_1,k,p,s_\perp)$  and
$Y^{\prime \sigma}_{\alpha\beta} (k_1,k,p,s_\perp)$.
Obviously, the gauge invariance of Eq. (\ref{www}) is not manifest.

Since we work  up to twist three, we  need to expand the
hard part  for the lowest-order diagram to the first
nonleading term. As for those shown in fig. 2, it is enough to
take their leading contributions.
Making use of the Ward identity, the twist-three contributions
associated with  the first derivative term in
the expansion of the lowest-order hard part can be
combined with  the leading  contributions  of  the
two diagrams  shown in fig. 2.  The net effect of such
a combination   is to replace the gluon field tensor
by the covariant derivative operator in the  two-variable
fragmentation matrices coupled  with  the hard parts of
the diagrams in fig. 2.

However, the Ellis-Furmanski-Petronzio scheme is not
a satisfactory procedure to extract
nonleading  twist  contributions \cite{qiu}.
The basic reason is that the leading term in the collinear
expansion contains as well the nonleading contributions.
To isolate such nonleading  effects implicit in the leading
term,  the efficient way  is to pull down a certain number
of ``special'' propagators into the hard part.
In our  twist-three case,  we need to
pull down $one$ special propagator  along with the
connected quark-gluon vertex  into the hard part,
either on the  left-hand side or on the right-hand side
of the final-state cut.   In other words,  twist-three
contributions hidden in the leading term of the  collinear
expansion of the lowest-order diagram  can
be taken into account by  including the two diagrams in fig. 3.
See what follows.

The notation of
the special propagator  \cite{qiu} can be introduced as follows.
Consider  the momentum carried by the quark propagator
in fig. 2 that links the electromagnetic vertex to the
quark-gluon one, one can parameterize it as
\begin{equation}
k^\mu= \hat k^\mu +\frac{k^2}{2k\cdot n} n^\mu,
\end{equation}
where
\begin{equation}
\hat k^\mu=\frac{1}{z} P^\mu +k^\mu_T -\frac{k^2_T}{2k\cdot n} n^\mu
\end{equation}
is the on-shell part  of $k^\mu$. Correspondingly,
the quark propagator  is decomposed into two parts:
\begin{equation}
\frac{i\rlap/k}{k^2+i\varepsilon}=
\frac{i\rlap/{\hat k}}{k^2+i\varepsilon }
+\frac{i\rlap/n}{2k\cdot n}.
\label{special}\end{equation}
Technically, $  i n\cdot \gamma /(2k\cdot n)$
is termed the special propagator by Qiu and labelled by adding a
bar on the normal propagator in the graphic representation.
To unravel the physical content of the special propagator,
let us consider the Fourier transformation
\begin{equation}
\int \frac{ dk^-}{2\pi}
\exp [-ik^-(\xi_1-\xi_2)^+]\frac{i\rlap/k}{k^2+i\varepsilon}
=\delta(\xi_1^+ - \xi^+_2) \frac{i\gamma^+}{2 k^+}
+\theta (\xi_1^+ -\xi^+_2) \frac{\rlap/{\hat k}}{2k^+}
\exp[-i\frac{k^2_T}{2 k^+}(\xi_1^+ -\xi^+_2)],
\end{equation}
from which  it can be seen that the
special propagator describes a  short-distance, or
``contact'' interaction in the light-cone.
Hence, as one goes beyond the leading twist
for the lowest-order diagram,
a special  propagator  along with the
connected quark-gluon vertex  should be
pulled down into the hard part.  Considering
that the diagrams shown in fig. 3 can be obtained
from the lowest-order diagram by pulling out a
quark propagator and its connected quark-gluon vertex,
it can also  be stated that the non-contact part of its linking
propagator should be  discarded,
because they have been included in $T_{\alpha\beta}
(k,p,s_\perp)$. It should be stressed that Qiu \cite{qiu}
 has demonstrated that
such a  cure can naturally reserve the  gauge invariance
for the hard part.

{}From the above discussions, we know that
the formula  for calculating the hadronic tensor reads
\begin{eqnarray}
\hat W^{\mu\nu}(q,p,s_\perp)&=& \frac{1}{4\pi N}
\int \frac{dz}{z}
{\rm Tr_D} {\rm Tr_C} \left[
H^{\mu\nu}_{(1)}(q,P/z)T(z,p,s_\perp)\right]
\nonumber \\
& &
+\frac{1}{4\pi N}\int d(\frac{1}{z_1}) dz
{\rm Tr_D} {\rm Tr_C}\left[\left(
(H^{\mu\nu\sigma}_{(2a)}(q,P/z,P/z_1)
+
H^{\mu\nu\sigma}_{(3a);{\rm spec}}(q,P/z)
\right)X_\sigma(z,z_1,p,s_\perp) \right]
\nonumber \\
& &
+\frac{1}{4\pi N}
\int d(\frac{1}{z_1}) dz
{\rm Tr_D} {\rm Tr_C}\left[\left(
H^{\mu\nu\sigma}_{(2b)}(q,P/z,P/z_1)
+H^{\mu\nu\sigma}_{(3b);{\rm spec}}(q,P/z)
\right)Y_\sigma(z,z_1,p,s_\perp)\right],
\label{www1}
\end{eqnarray}
where
\begin{equation}
T_{\alpha\beta} (z,p,s_\perp)=
z \sum\limits_X\int  \frac{d\lambda }{2 \pi}
\exp ( -i\lambda/z)
\langle0|\psi_{\alpha} (0)|\Lambda (p,s_\perp),X\rangle
\langle\Lambda (p,s_\perp),X|\bar \psi_{\beta} (\lambda n )|0\rangle,
\label{m1}
\end{equation}
\begin{eqnarray}
X^\sigma_{\alpha\beta} (z_1,z,p,s_\perp)&=&
\sum\limits_X\int \frac{ d\lambda_1d   \lambda }{(2\pi)^2}
\exp \left(-i\lambda_1 (1/z-1/z_1)-i\lambda/z\right) \nonumber  \\
&  & \times
\langle0| {D}^\sigma (\lambda_1   n)\psi_\alpha (0)
|\Lambda (p,s_\perp),X\rangle\langle\Lambda (p,s_\perp),X|
\bar\psi_\beta (\lambda   n) |0\rangle,
\label{u}\end{eqnarray}

\begin{eqnarray}
Y^\sigma_{\alpha\beta} (z_1,z,p,s_\perp)&=&
\sum\limits_X\int \frac{ d\lambda_1d   \lambda }{(2\pi)^2}
\exp (-i\lambda (1/z_1-1/z)-i\lambda_1/z_1) \nonumber  \\
&  &\times
\langle0|
\psi_\alpha (0)|\Lambda (p,s_\perp),X\rangle\langle\Lambda (p,s_\perp),X|
\bar\psi_\beta (\lambda_1   n) \stackrel{\leftarrow}{D^\sigma}
(\lambda   n)|0\rangle.
\label{v}\end{eqnarray}
In our work, the covariant derivative operator is defined as
$D^\sigma=i\partial^\sigma -gA^\sigma$ and
$\stackrel{\leftarrow}{D^\sigma}=i\stackrel{\leftarrow}{\partial^\sigma}
 -gA^\sigma$.

 To arrive at  factorized expressions,  we decompose
the above  three  fragmentation matrices in the
Dirac and Lorentz  spaces.
Again, we suppress  the  spin-independent
terms as well as those arising from the final-state interactions
for the brevity of formulas. As a result,

\begin{eqnarray}
T_{\alpha\beta}(z,p,s_\perp)
&=& [\hat h_1 (z) \gamma_5 \rlap/s_\perp \rlap/P
+M\hat g_T (z) \gamma_5 \rlap/s_\perp+\cdots]_{\alpha\beta}, \label{dec1}
\end{eqnarray}

\begin{equation}
X^\sigma_{\alpha\beta}(z_1,z,p,s_\perp)=
\frac{iM}{2z}\hat G_1 (z_1,z)
\varepsilon_{\sigma \rho\tau\eta}s^\rho_\perp P^\tau n^\eta
\rlap/P_{\alpha\beta}
+
\frac{M}{2z}\hat G_2(z_1,z) s^\sigma_\perp (\gamma_5 \rlap/P)_{\alpha\beta}
+\cdots ,
\label{udec}
\end{equation}
\begin{equation}
Y^\sigma_{\alpha\beta} (z_1,z,p,s_\perp)= -\frac{iM}{2z}\hat G_1 (z_1,z)
\varepsilon_{\sigma \rho\tau\eta}s^\rho_\perp P^\tau n^\eta
\rlap/P_{\alpha\beta}
+
\frac{M}{2z}\hat G_2 (z_1,z) s^\sigma_\perp (\gamma_5 \rlap/P)_{\alpha\beta}
+\cdots ,
\label{vdec}
\end{equation}
where $\hat h_1 (z) $, $\hat g_T(z) $, $\hat G_1(z_1, z) $,  and
$\hat G_2(z_1, z) $
are Jaffe and Ji's parton  fragmentation matrix elements, whose definitions
can be easily  projected out from the above decompositions.

 Using QCD equation of motion, one can prove \cite{Ji}  that
\begin{equation}
\int d(\frac{1}{z_1}) [ \hat G_1 (z_1,z) + \hat G_2 (z_1,z)] =
-\frac{1}{z}\hat g_T (z)+ \frac{m}{M}  \hat h_1 (z),  \label{proj}
\end{equation}
where $m$ is the quark mass.

Inserting Eqs. (\ref{dec1})--(\ref{vdec}) into   (\ref{www1}) and
completing the algebra, we obtain
\begin{eqnarray}
\hat W_{\mu\nu}(q,p,s_\perp)&=&
\frac{i}{z_0^2 (P\cdot q)}
\sum\limits_f e_f^2
\left[z_0m_f \hat h^f_1(z_0)+ M \hat g^f_T(z_0)\right]
\varepsilon_{\mu\nu\tau\rho}q^\tau s^\rho_\perp.
\label{res}
\end{eqnarray}
Here we note this
result is complete up to twist three.
Again, we make substitutions $p\to P$ and $z_0 \to z_B$.
By confronting  with  Eq. (\ref{dec}),
we  attain for  $\hat g_1$ and $g_2$
the following relation:
\begin{equation}
\hat g_1(x_B) +\hat g_2(x_B) =\frac{1}{z^2_B}\sum\limits_f e_f^2
\left[\frac{m_f}{M}z_B  \hat h^f_1(z_B)+ \hat g^f_T(z_B)\right]. \label{ss2}
\end{equation}

\section{conclusion}

We conclude this paper  by  discussing the phenomenological
implications of  our factorization results about $\hat g_1$
and $\hat g_2$.  Here we assume that the $\Lambda$ hyperon
is predominantly produced via the strange quark fragmentation.
Then, both  in Eq. (\ref{ss1}) and  in (\ref{ss2}) the flavor summation
can be dropped.  Futhermore,  the terms associated with
the quark mass in (\ref{ss2}) can be  ignored at a first-order
approximation because  $m_s/M\approx 0.1$.
Although not discussed here, we  can believe that by controlling
judiciously the polarization of the initial-state electron
beam and analyzing the  final-state $\Lambda$ polarization,
it is possible to measure $\hat g_1$ and $\hat g_2$.
Therefore,  the measurement of these two structure functions
will  allow for the determination of  two $s\to \Lambda$
fragmentation functions $\hat {\sl g}_1 (z)$ and $\hat g_T(z)$.

  It should be noted that
Eq. (\ref{ss1}) might be  not so helpful.
The reason is that $\hat {\sl g_1} (z)$
can be  independently measured at the  LEP, which Burkardt
and  Jaffe have discussed in Ref. \cite{cite1}.
At  the $Z^0$ resonance,  the parity violation
can naturally  generate longitudinally-polarized
strange quarks so  one does not need to polarize the
initial-state beam.  However, Eq. (\ref{ss2}) is by no means  trivial.
So far,  the measurement of chiral-odd nonperturbative
matrix elements, to which $\hat h_1(z)$ and $\hat g_T$  belong,
 have been a challenge to the particle
physics community.  So far, the  general strategy to
treat them has been tackling them in pairs [1],  [2],
\cite{levelt}. Our results, Eq. (\ref{ss2}), indicates that
once we have known $\hat h_1(z)$ from other sources
or it can be neglected at the first approximation,
the measurement of $\hat g_1$ and $\hat g_2$,
present us a  relatively clean situation,
in which  information about twist-three fragmentation
function $\hat g_T(z)$ can be obtained.
The penalty in such a  scheme is to polarize one of the
initial-state beams. As is well known,  it is now
quite feasible to  obtain highly polarized electron beams.

  In summary, we  worked out a QCD field theory
study  of $\hat g_1$ and $\hat g_2$, two spin-dependent
structure functions for the inclusive spin-half baryon
production in electron-positron annihilation.
Making use of the collinear expansion procedure
by Ellis, Furmanski and Petronzio, and the special
propagator concept by Qiu,  we  derived a formula
relating the sum of $\hat g_1$ and $\hat g_2$ to
the transverse-spin-dependent
quark fragmentation  functions $\hat h_1(z)$ and  $\hat g_T(z)$.
On the basis of this finding, we pointed  out  that
it is possible to  extract data about $\hat g_T(z)$
in the inclusive $\Lambda$ hyperon production by
electron-positron annihilation.
Considering the common believe that chiral-odd  hadron matrix
elements and/or parton fragmentation matrix elements  should
be measured in pairs, our   results can be taken as an interesting
counter example.

One of the authors (W. L.)  thanks  Professors E. Leader  and X. Ji
for  providing  some of their papers before publication.

\newpage
\pagestyle{empty}

\centerline{Figure Captions}

\begin{enumerate}

\item{The lowest-order cut diagram for the inclusive
$\Lambda$ hyperon production by a time-like photon.}

\item{The cut diagram for the inclusive
hyperon production by a time-like photon
with one  gluon correlation at the stage of parton fragmentation.}

\item{The cut diagram for the inclusive
$\Lambda$ hyperon production by a time-like photon
with  one  gluon radiation in the quark fragmentation.
One ``special'' propagator is pulled down into the hard partonic interaction
part, either (a) on the left-hand side or (b) on the right-hand side
of the final-state cut.}

\end{enumerate}

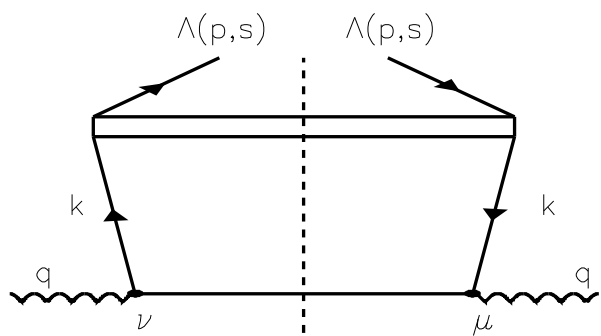
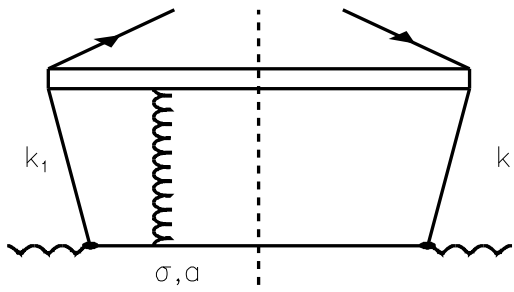
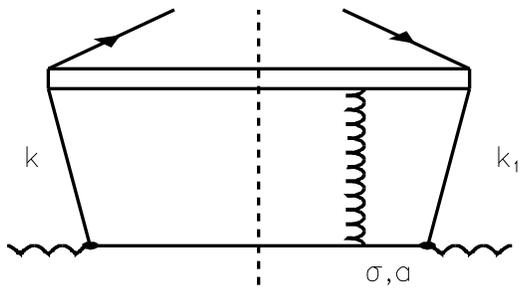
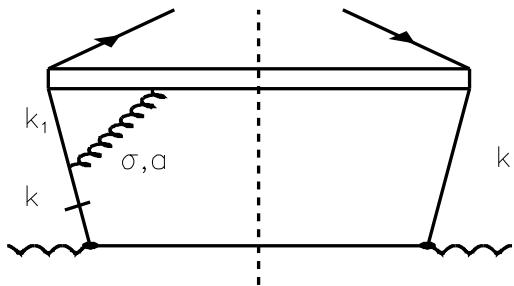
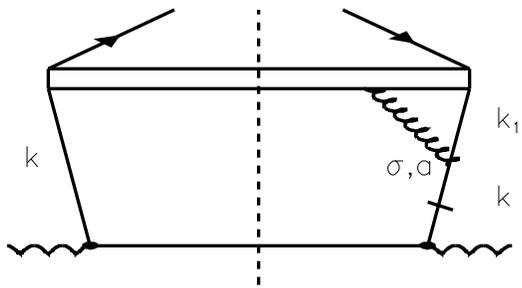
\end{document}